\documentclass[prb,twocolumn,english,superscriptaddress,floatfix]{revtex4}
\usepackage{graphicx}
\usepackage{bm, amsmath, amssymb}
\usepackage{pdfpages}
\usepackage[T1]{fontenc}
\usepackage[latin9]{inputenc}
\usepackage{amstext}
\usepackage{esint}
\newcommand{\nbar}[0]{\bar{n}}

\newcommand{\pii}[0]{\rho_{11}}
\newcommand{\poi}[0]{\rho_{01}}
\newcommand{\poo}[0]{\rho_{00}}

\def\be{\begin{equation}}
\def\ee{\end{equation}}
\def\bea{\begin{eqnarray}}
\def\eea{\end{eqnarray}}

\usepackage{babel}

\begin{document}

\title{Mapping the optimal route between two quantum states}

\author{S. J. Weber}
\affiliation{Quantum Nanoelectronics Laboratory, Department of Physics, University of California, Berkeley CA 94720}
\author{A. Chantasri}
\affiliation{Department of Physics and Astronomy and Center for Coherence and Quantum Optics, University of Rochester, Rochester, New York 14627}
\author{J. Dressel}
\affiliation{Department of Electrical Engineering, University of California, Riverside, CA 92521}
\author{A. N. Jordan}
\affiliation{Department of Physics and Astronomy and Center for Coherence and Quantum Optics, University of Rochester, Rochester, New York 14627}
\affiliation{Institute for Quantum Studies, Chapman University, University Drive, Orange, California 92866}
\author{K. W. Murch}
\affiliation{Department of Physics, Washington University, St.\ Louis, Missouri 63130}
\author{I. Siddiqi}
\affiliation{Quantum Nanoelectronics Laboratory, Department of Physics, University of California, Berkeley CA 94720}

\date{\today}

\maketitle

{\bf
A central feature of quantum mechanics is that a measurement is intrinsically probabilistic.  As a result,  continuously monitoring a quantum system will randomly perturb its natural unitary evolution.  The ability to control a quantum system in the  presence of these fluctuations is of increasing importance in quantum information processing and finds application in fields ranging from nuclear magnetic resonance\cite{vand05} to chemical synthesis\cite{shap11}. A detailed understanding of this stochastic evolution is essential for the development of optimized control methods.   Here we reconstruct the individual quantum trajectories\cite{carm93, carm94,wisebook} of a superconducting circuit that evolves in competition between continuous weak measurement and driven unitary evolution.  By tracking individual trajectories that evolve between an arbitrary choice of initial and final states we can deduce the most probable path through quantum state space. These pre- and post-selected quantum trajectories also reveal the optimal detector signal in the form of a smooth time-continuous function that connects the desired boundary conditions. Our investigation reveals the rich interplay between measurement dynamics, typically associated with wave function collapse, and unitary evolution of the quantum state as described by the Schr\"odinger equation.  These results and the underlying theory\cite{chan13}, based on a principle of least action, reveal the optimal route from initial to final states, and may enable new quantum control methods for state steering and information processing. }

Our experiment focuses on the dynamics of two quantum levels of a superconducting circuit (a qubit), which can be continuously measured and excited by microwave pulses.   To access individual quantum trajectories, we make use of the fact that fully projective measurement (or wavefunction collapse) happens over an average timescale $\tau$ controlled by the interaction strength between the system and the detector.  By recording the measurement signal in time steps much shorter than $\tau$ with high fidelity, we realize a continuous sequence of weak measurements and track the qubit state as it evolves in a single experimental iteration.  Individual weak measurements have been recently employed in atomic physics experiments that probe wave function collapse \cite{guer07} and perform state stabilization \cite{sayr11}.  In the domain of superconducting circuits, weak measurements\cite{hatr13} have only recently been realized due to the challenge associated with high fidelity detection of near single-photon level microwave signals.  Advances in superconducting parametric amplifiers have enabled continuous feedback control \cite{vija12,blok13,lang13}, the observation of individual quantum trajectories \cite{murc13traj,jord13}, the determination of weak values \cite{groe13,camp13}, and entanglement of qubits\cite{rist13,roch14}.

\begin{figure}
\includegraphics[angle = 0, width = .5\textwidth]{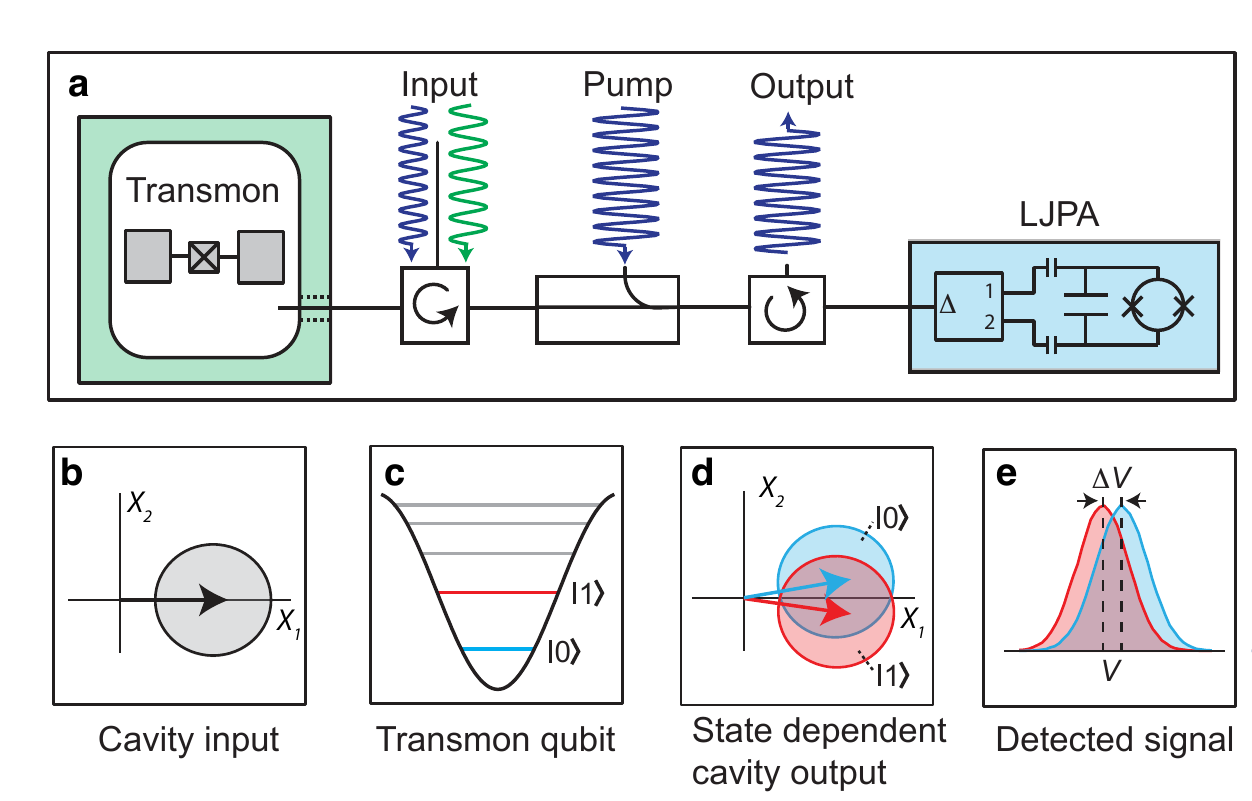}
\caption{\label{fig1} Setup. ({\bf a}) A transmon circuit is dispersively coupled to a three dimensional copper waveguide cavity.  Microwave signals that reflect off the cavity port are amplified by a Lumped-element Josephson Parametric Amplifier (LJPA) \cite{hatr11para} operating near the quantum limit. ({\bf b}) A microwave tone that probes the cavity near resonance is shown as a phasor in the $X_1$-$X_2$ plane with zero-point quantum fluctuations shown by the shaded region.  ({\bf c}) Ground and excited energy levels are shown on the transmon potential. ({\bf d}) The reflected microwave tone acquires a qubit state-dependent phase shift that is smaller than the quantum fluctuations of the measurement signal.  After further amplification, the $X_2$ quadrature of the measurement tone is digitized. ({\bf e}) The measurement is calibrated by examining the distributions of measurement signals for the qubit prepared in the $|0\rangle$ (blue) and $|1\rangle$ (red) states.
}
\end{figure}

Our experiment consists of a superconducting transmon circuit \cite{koch07} dispersively coupled to a waveguide cavity \cite{paik113D} (Fig.\ 1a).  Considering only the two lowest levels of the transmon as a qubit, our system is described by the Hamiltonian $H = H_{0}+ H_\mathrm{int} + H_\mathrm{R}$,
\begin{eqnarray}  \label{H}
H_\mathrm{int} = -\hbar \chi a^\dagger a \sigma_z, \  H_\mathrm{R} = \hbar\frac{\Omega}{2} \sigma_y,
\end{eqnarray}
where  $H_0$ describes the qubit and cavity, $\hbar$ is the reduced Plank's constant,  $a^\dagger (a)$ is the creation (annihilation) operator for the cavity mode, and $\sigma_{y,z}$ are qubit Pauli operators.  $H_\mathrm{R}$ describes a microwave drive at the qubit transition frequency which induces unitary evolution of the qubit state characterized by the Rabi frequency $\Omega$.  $H_\mathrm{int}$ is the interaction term, characterized by the dispersive coupling rate $ \chi/2\pi=-0.6$ MHz.  This term describes a qubit state-dependent frequency shift of the cavity which we use to perform quantum state measurement in our system.  As depicted in Figure 1b-e, a microwave tone that probes the cavity near its resonance frequency will acquire a qubit state-dependent phase shift.  If the measurement tone is very weak, quantum fluctuations of the electromagnetic mode fundamentally obscure this phase shift, resulting in a partial or weak measurement of the qubit state.  We use a near-quantum-limited parametric amplifier \cite{cast08, hatr11para}  to amplify the $X_2$ quadrature of the reflected signal, which is proportional to the qubit state-dependent phase shift.    After further amplification, we digitize the signal  in 16 ns time steps  resulting in a measurement signal $V(t)$.  
Each time step is small compared to the characteristic measurement time, $\tau = \kappa/(16  \chi^2 \nbar \, \eta_\mathrm{col} \eta_\mathrm{amp})$, where $\nbar$ is the average intracavity photon number,  $\kappa/2\pi =  9.0$ MHz is the cavity decay rate, and $\eta_\mathrm{col}\eta_\mathrm{amp}$ is the measurement quantum efficiency \cite{koro11} that decomposes into separate collection and amplification efficiencies. The characteristic measurement time is calibrated by examining (Gaussian) histograms of the measurement results for the qubit prepared in the  $\sigma_z$ eigenstates  $|0\rangle$ and $|1\rangle$ and is given by the time it takes to separate the two distributions by two standard deviations, $\Delta V = 2 \sigma$.  

   \begin{figure*}\begin{center}
\includegraphics[angle = 0, width = .85\textwidth]{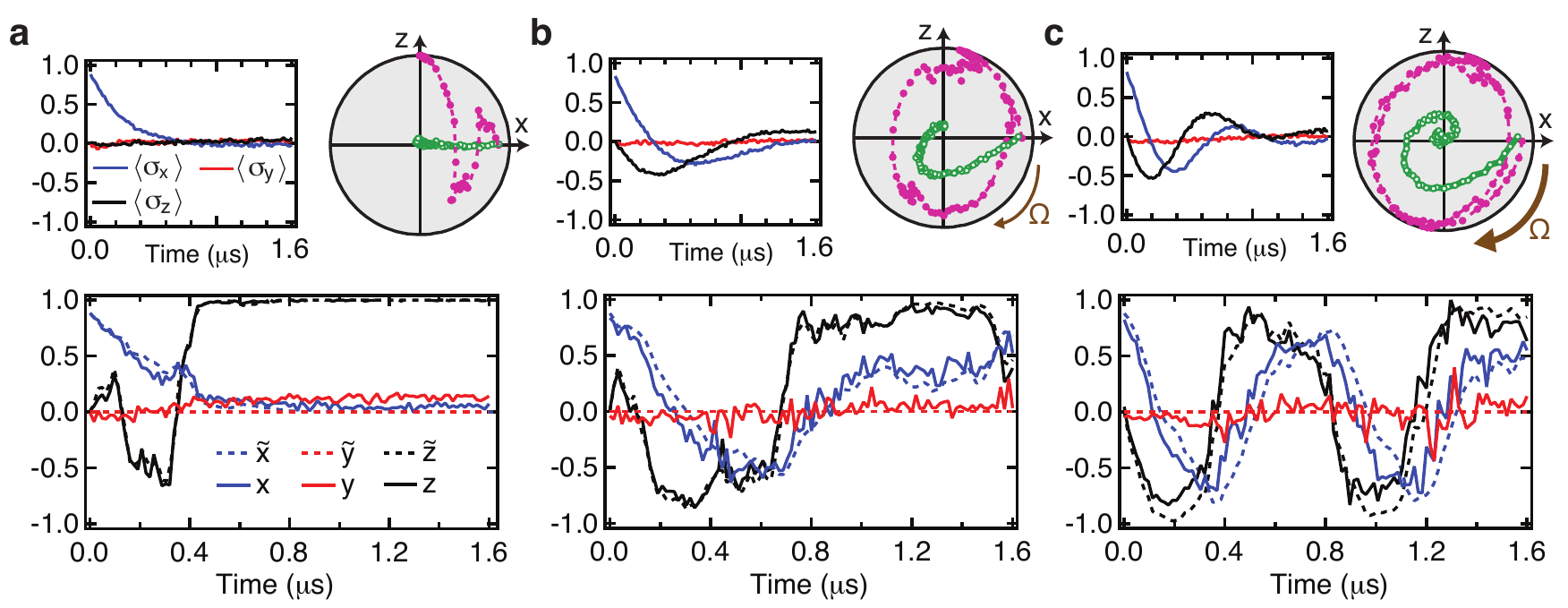}
\end{center}
\caption{ \label{fig2}  Quantum trajectories of the quantum state $x$ (blue), $y$ (red), and $z$ (black) are plotted versus time.  The upper panels depict the full ensemble evolution sided by individual trajectories (magenta) and the ensemble average (green) plotted in the $ x$--$ z$ plane of the Bloch sphere. The lower panels depict individual quantum trajectories (dashed curves), with comparison to their tomographic reconstructions (solid curves). Panels ({\bf a}), ({\bf b}), ({\bf c}) correspond to different values of Rabi drives, $\Omega/2\pi$ = 0, 0.56, and 1.08 MHz, respectively.  Here, $\tau =315$ ns and $\Gamma = 3.85 \times 10^6$ s$^{-1}$.}
\end{figure*}

\begin{figure}
\includegraphics[angle = 0, width = .45\textwidth]{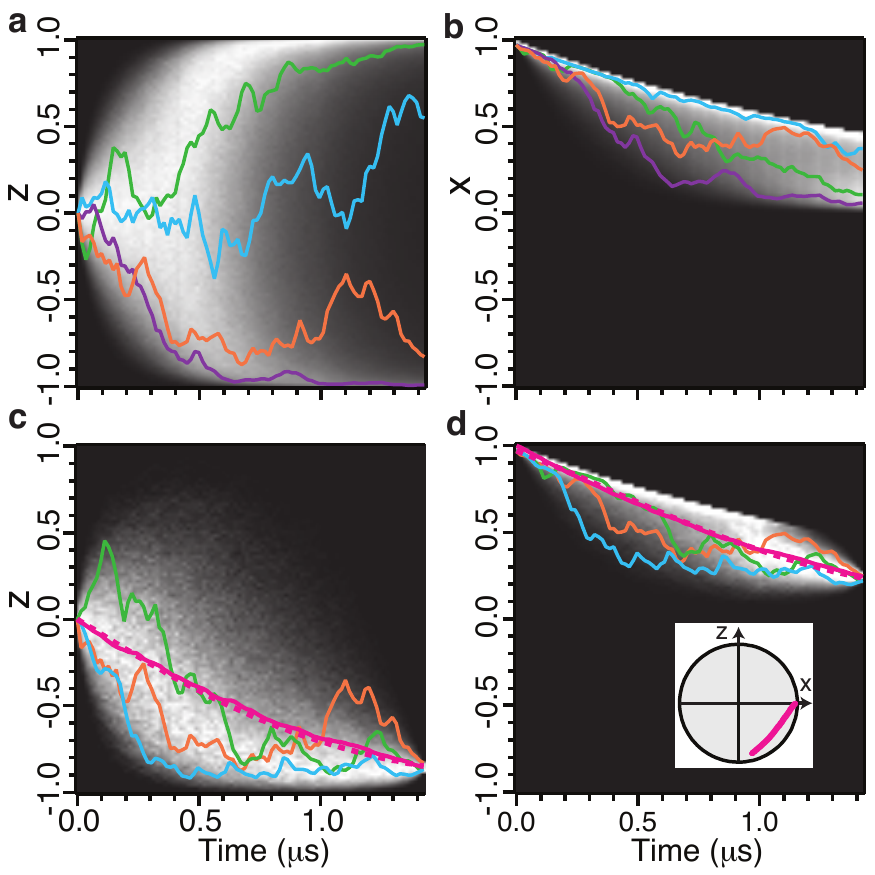}
\caption{\label{fig3}  Greyscale histograms of quantum trajectories in the undriven case for a measurement duration of 1.424 $\mu s$.  ({\bf a, b}) Histogram of all measured $z$, and $x$ trajectories respectively, beginning from state $(x_I=0.97, z_I=0)$.  Representative trajectories are shown in color.  ({\bf c, d})  Histogram of trajectories $z$, $x$ respectively, conforming to the final chosen boundary condition, $z_F = -0.85 \pm .03$.  Magenta curves are most likely trajectories for the experimental data (solid) and from the theory (dashed).  Representative trajectories are shown in other colors.  Here, $\tau = 1.25\ \mu$s,  and $\Gamma = 0.94\times10^6$ s$^{-1}$.}
\end{figure}

\begin{figure*}
\includegraphics[angle = 0, width = 1\textwidth]{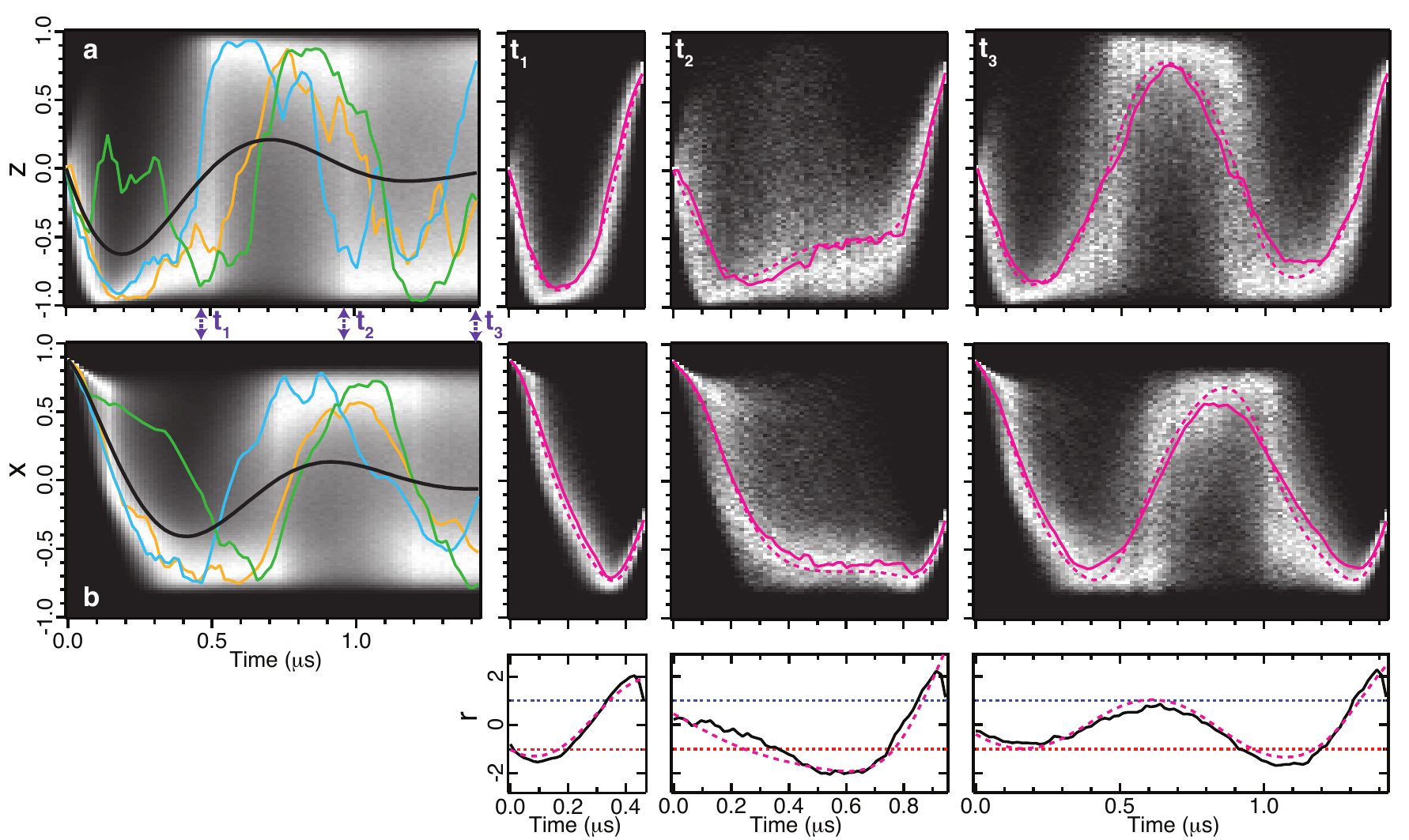}
\caption{\label{fig4}   Greyscale histograms of quantum trajectories in the driven case beginning from state $(x_I=0.88, z_I=0)$.  Here, $\tau = 315$ ns, $\Gamma = 3.85 \times 10^6$ s$^{-1}$, $\Omega/2\pi$ = 1.08 MHz.   Panels ({\bf a}) and ({\bf b}) show histograms with representative trajectories plotted in color with the average trajectory shown in black.   In the other panels, we post-select on the final state $(z_F = 0.7, x_F = -0.29)$, with a post-selection window of $\pm 0.08$.  Post-selected histograms for $z$ are given in the top panels, and $x$ histograms are given in the bottom panels.  Magenta curves are most likely trajectories for the experimental data (solid) and from the theory (dashed).  As the time duration between the boundary conditions is increased from $t_1 =0 .464\ \mu s$ to $t_2 = 0.944\ \mu s$, and $t_3 = 1.424\  \mu s$, the most likely trajectory connecting the initial and final states changes dramatically but is well described by the theory (dashed).  The bottom panels compare the optimal detector signals ($r$), shown as dashed lines to the conditioned average signal (weak functions), black lines, for each case.}
\end{figure*}

In our experiment, we prepare the qubit along the $x$ axis of the Bloch sphere by heralding the $|0\rangle$ state and applying a $\pi/2$ rotation about the $y$ axis. Then, a measurement tone at 6.8316 GHz continuously probes the cavity for a variable time $t$, which weakly measures the qubit in the $\sigma_z$ basis. Finally, we apply further rotations and perform a projective measurement to conduct quantum state tomography.   In figure 2 (top panels) we show the ensemble averaged tomography for $\Omega/2\pi$ = 0, 0.56, and 1.08 MHz.  From these curves, we extract $\Omega$ and the ensemble decay rate $\Gamma$, from which we calculate a total quantum efficiency $\eta_\mathrm{tot}= 1/(2\, \tau\, \Gamma) =\eta_\mathrm{col}\eta_\mathrm{amp}\eta_\mathrm{env} = 0.4$ where the last factor indicates the (nearly negligible) extra environmental dephasing $\eta_\mathrm{env}= (1+ \kappa/8 \chi^2 \nbar T^*_2)^{-1}$ with $T^*_2 = 15\mu$s.

In each iteration of the experiment, we can use the recorded measurement signal to calculate the best estimate for the qubit state conditioned on the measurement record.  As discussed in the supplementary information, at each time-step we apply a two step update procedure to track the evolution of the system density matrix $\rho$.  We account for the measurement result using a quantum generalization of Bayes' rule \cite{koro99,koro11}, and for the Rabi drive by applying a unitary rotation. Our finite detector efficiency reflects an imperfect knowledge about the state of the system and results in a decay of coherence given by rate $\gamma = \Gamma - 1/(2 \tau) $.  From the density matrix $\rho$, we calculate expectation values of the Pauli operators conditioned on the measurement signal,  $x \equiv \mathrm{Tr}[\rho \sigma_x]$, $y \equiv \mathrm{Tr}[\rho \sigma_y]$, and $z \equiv \mathrm{Tr}[\rho \sigma_z]$.

In figure 2a  we display a sample trajectory with no drive ($\Omega$ = 0) that shows the stochastic motion of the qubit state as it evolves under measurement, and is ultimately projected into the $|0\rangle$ state. As described in the methods, we use conditioned quantum state tomography to  reconstruct  the trajectory. Figure 2b,c demonstrate that we can track the state faithfully in the presence of unitary state evolution induced by a drive at the qubit frequency.  These trajectories highlight the stark difference between ensemble dynamics and the dynamics of individual quantum trajectories; while the ensemble decays rapidly to a mixed state, the individual trajectories remain remarkably pure despite the modest quantum efficiency of $\eta_\mathrm{tot} = 0.4$.

Using this ability to track individual trajectories starting from a given initial state, we now consider the sub-ensemble of trajectories that arrive at a particular final state at a given time.
 This sub-ensemble allows us to examine the conditional quantum dynamics of the state  that satisfy two boundary conditions, one in the past, ``pre-selection'', and one in the future, ``post-selection''. This is similar to an analysis that leads to weak values\cite{wana55,ahar64,ahar88}, and time continuous generalizations\cite{will08,camp13} which consider an additional projective post-selection measurement. In contrast to that approach, we use only a single continuous measurement: the pre-selection is just the initial state, and the post-selection is simply what the state is when the detector stops measuring.   The resulting average of the measurement output gives a ``weak function"  that connects the boundary conditions.

To investigate the full ensemble and post-selected sub-ensemble dynamics, we perform $10^{5}$ iterations of the experiment with a measurement duration of 1.424 $\mu$s.  For each experiment, we construct the quantum state trajectory by finding $x$ and $z$ for every time step.  Figure 3 displays the measurement dynamics for $\Omega= 0$.  We consider the sub-ensemble of trajectories that had final values $z(1.424\  \mu\mathrm{s}) = -0.85\pm 0.03$ and $x(1.424\  \mu\mathrm{s}) =0.23\pm 0.03$.   This analysis allows us to examine properties of the conditional trajectories such as the most likely path that connects pre- and post-selected states.   

The most likely paths  can be theoretically calculated based on a stochastic path integral representation of the joint probability of the measurement outcomes at every point in time with boundary condition constraints.    The conditional detector backaction on the quantum state can be imposed at every timestep with Lagrange multipliers $(p_x, p_y, p_z)$ as auxiliary dynamical parameters.  Finding the extremum of the stochastic action leads to equations of motion for the optimal path connecting  the boundary conditions.  As we discuss in supplementary information this corresponds to optimizing the total path probability between the states. Since the experiment operates in the $x$--$z$ plane of the Bloch sphere, the (deterministic) equations of motion for the optimized path are
\begin{subequations}\label{eoms}
\begin{eqnarray}  
\dot{x} &=& -\, \gamma \, x  + \Omega\, z - x z\, r /\tau,  \\
\dot{z} & =& -\, \Omega \, x + (1-z^2)\, r /\tau,  \\
\dot{p}_x &=& +\,\gamma \, p_x +\Omega\, p_z + p_x z \,r/\tau,  \\ 
\dot{p}_z &=& -\,\Omega\, p_x + (p_x x + 2 p_z z -1)\,r/\tau, 
\end{eqnarray}
\end{subequations}
where $x,z,p_x, p_z,r$ are now functions of time and $r=z+ p_z (1-z^2) -  p_x x z$. We note that $r$ may be interpreted as the most likely detector estimate of the qubit's $z$-coordinate, where the noisy fluctuations around $z$ have now been constrained by the Lagrange multipliers $p_x, p_z$. This readout relates to the optimal detector signal as $V_\mathrm{opt}  =\Delta V r/2$. The solution to these nonlinear equations admits four constants of motion, which permits the imposition of both initial $(x_I, z_I)$ and final $(x_F, z_F)$ boundary conditions.

The equations have a simple analytic solution $(\bar{x}, \bar{z})$ for $\Omega=0$.  We consider measurement for a time $T$, starting in the initial state $(x_I=1, z_I=0)$, and ending in a state $(x_F, z_F)$ (in this particular case, $x_F$ is determined by the choice of $z_F$). The solution of Eqs.~\eqref{eoms} is $\bar{x}(t) = \, e^{-\gamma t}\, \text{sech}\, \bar{r}\, t/\tau, \bar{z}(t) =  \,\tanh \bar{r}\, t/\tau$ where $\bar{r}=(\tau/T) \tanh^{-1}z_F$, is the detector output of maximum likelihood. These solutions are plotted in Figure 3 showing agreement with the experimentally obtained most likely path (see methods).

In figure~\ref{fig4}, we display the full ensembles and post-selected ensembles for the driven case ($\Omega/2\pi = 1.08$ MHz). Depending on the amount of time between the initial and final states, the competition between measurement and Schr\"odinger dynamics produces different (and nontrivial) optimal routes, showing alternatively diffusive Rabi oscillation dynamics or quantum jump dynamics (where the system is effectively pinned in one of the eigenstates)  \cite{vija11,gamb08,chan13,koro99}.  We compare the experimental most likely trajectories (see methods) to the most likely paths obtained from solving Eqs.~\eqref{eoms}.  The equations were numerically solved with a shooting method to satisfy both initial and final boundary conditions at different times. These numerical solutions show reasonable agreement with the experimental most likely curves.

In addition to the quantum paths, the solution of Eqs.~\eqref{eoms} also gives the optimal detector response to move the quantum system to the target state after a given time. We compare these optimal signals to the conditioned average detector signals (weak functions) in Figure 4. 
 The post-selection allows the conditioned average detector signal to exceed the usual range of $[-1,1]$ for $z$. This behavior is analogous to that of weak values which can also lie outside their eigenvalue range\cite{ahar88}.

The ability to find and verify the most likely path between chosen initial and final quantum states under continuous measurement advances the field of quantum control of individual systems and our fundamental understanding of quantum measurement.  
Our detailed experimental and theoretical investigation of the probability distributions for mapping quantum paths allows for future optimization of control parameters to tune the most-likely behavior in quantum systems. 

\small
{\bf Methods}
Details about the sample fabrication, experimental setup, and data analysis are given in the supplemental information.

To verify that we have accurately tracked the quantum state of the system, we perform quantum state tomography at discrete times along the trajectory. We denote the target trajectory, which is based on a single run of the experiment, $\tilde{x}(t), \ \tilde{z}(t)$. For each experimental sequence of total measurement duration $t$ we propagate $\rho$ and if $x(t) = \tilde{x}(t) \pm 0.03 $ and $z(t) = \tilde{z}(t)\pm 0.03$, the subsequent tomography results are included in the tomographic reconstruction of the state at time $t$.  We repeat this analysis for all time steps between 0 and 1.6 $\mu$s, showing good agreement between the individual trajectories and the tomographic reconstructions. 

The experimental most likely path was obtained by computing a path probability for each trajectory and then averaging the trajectories in the top 5-8\%.  This 5-8\% window was motivated by an independent  Monte Carlo simulation with the same sample size.  In the supplemental information we discuss the detailed estimation of the experimental most likely path and its correspondence with the solutions of Eqs.~\eqref{eoms}.

{\bf Acknowledgements}  We thank A. N. Korotkov, S. G. Rajeev,  N. Roch, and D. Toyli for discussions. This research was supported in part by the Army Research Office, Office of Naval Research and the Office of the Director of National Intelligence (ODNI), Intelligence Advanced Research Projects Activity (IARPA), through the Army Research Office. All statements of fact, opinion or conclusions contained herein are those of the authors and should not be construed as representing the official views or policies of IARPA, the ODNI or the US government. A.N.J acknowledges support from NSF grant no. DMR-0844899 (CAREER)

%{\bf Supplementary Information} is linked to the online version of the paper at www.nature.com/nature.

%{\bf Author contributions}
%S.J.W. and K.W.M. performed the experiment and analyzed the experimental data. J.D and A.C. wrote the trajectory simulation code. A.C., J.D. and A.N.J. contributed the theory. All work was carried out under the supervision of I.S.  All authors contributed to writing the manuscript.

%{\bf Author Information} 
% Reprints and permissions information is available at www.nature.com/reprints.
 %The authors declare no competing financial interests. 
 \vspace {.1in}
 
Correspondence and requests for materials should be addressed to K.W.M. (murch@physics.wustl.edu)

%\bibliographystyle{naturemag}
%\bibliography{pre_and_post_references}

%\bibliographystyle{naturemag}
%\bibliography{weakrefs}
\pagebreak

\begin{widetext}

\begin{center}
%Supplementary information for:  \\ ``Mapping the optimal route between two quantum states''
{\bf SUPPLEMENTARY INFORMATION FOR: \\``MAPPING THE OPTIMAL ROUTE BETWEEN TWO QUANTUM STATES''}
\end{center}

\section{Experimental methods}

This section details information about device parameters, experimental setup and  data analysis routines. 

\subsection{Device parameters}

The qubit consists of two aluminum paddles connected by a double-angle-evaporated aluminum SQUID deposited on double-side-polished silicon.  The qubit is characterized by a charging energy $E_c/h = 200$ MHz, and a Josephson energy $E_J/h = 11$ GHz.  The qubit is operated with negligible flux threading the SQUID loop with a transition frequency $\omega_q/2\pi =  4.01057$ GHz.  The qubit is located off center of a 6.8316-GHz copper waveguide cavity.  With the measurement tone on the qubit transition frequency was ac-Stark shifted to 4.00748 GHz.  Qubit pulses and drive are performed at the ac-Stark shifted frequency.

The Lumped-element Josephson Parametric Amplifier (LJPA) consists of a two-junction SQUID, formed from 2-$\mu$A Josephson junctions shunted by 3 pF of capacitance and is flux biased to provide $20$ dB of gain at the cavity resonance frequency.  The LJPA is pumped by two sidebands equally spaced 300 MHz above and below the cavity resonance.  

\subsection{Experiment setup}

Figure S1 displays a schematic of the experimental setup.  Experimental sequences start with an 800-ns readout to herald the $|0\rangle$ state $(z = +1)$, followed by a 16-ns $\pi/2 $ rotation about the $y$ axis to prepare the qubit along the $x$ axis.  After a period of variable duration, we perform quantum state tomography by applying either  rotations about the $x$ and $y$ axes or no rotation followed by a second 800-ns readout.  Tomography results were corrected for the readout fidelity of $95\%$.

\subsection{Calibration of the measurement}

We calibrate the characteristic measurement time $\tau$ by examining histograms of the measurement signal for the qubit prepared in either the $|0\rangle$ or $|1\rangle$ states.  We prepare these states through a herald readout and then digitize the measurement signal for a variable period of time.  The resulting distributions are approximately Gaussian,
 \begin{eqnarray}
P(V \, |\, 0) =\sqrt{\frac{ 1}{2 \pi \sigma^2}}\, e^{-\frac{ 1}{2 \sigma^2}(V-\Delta V/2)^2},\\
P(V \, |\, 1) =\sqrt{\frac{ 1}{2 \pi \sigma^2}}\, e^{-\frac{ 1}{2 \sigma^2}(V+\Delta V/2)^2}.
\end{eqnarray}
 We fit the distributions to determine $\Delta V$, the voltage separation of the peaks, and the variance $\sigma^2$.  The quantity $S =\Delta V^2/\sigma^2$ increases linearly with integration time, $S = 4 t/\tau$, which we fit to determine  the characteristic measurement  time $\tau$.  

To calibrate the initial state and total dephasing rate, we prepare the qubit along the $x$ axis and perform quantum state tomography after a variable period of time. The tomography results for the full ensemble are shown in Figure 2a (in the main text), and exhibit exponential decay of coherence at rate $\Gamma$.  The total quantum measurement efficiency is given by $\eta_\mathrm{tot} = 1/(2\Gamma \tau)$.  Note that the total quantum measurement efficiency $\eta_\mathrm{tot} = \eta_\mathrm{col}\times \eta_\mathrm{amp}\times \eta_\mathrm{env}$, is the product of efficiencies for collection, for amplification and from extra environmental dephasing. We use the tomography value at $t=0$ to determine the initial state, denoted $x_0,\ z_0$. 

To determine the Rabi frequency, $\Omega$, we examine the ensemble tomography results as shown in Figure 2b,c (in the main text).  The ensemble evolution is given by the Lindblad equation with arbitrary Rabi drive, $\dot{x}(t) = -\Gamma x(t) + \Omega z(t), \ \dot{z}(t) = -\Omega x(t)$.  With initial state $x_0$ and $z_0$, these equations have an analytic solution, 
\begin{eqnarray}
x(t) = e^{-\Gamma t/2} \left( x_0 \mathrm{cos }\ \lambda t - \frac{ \Gamma x_0-2 \Omega z_0}{2 \lambda} \mathrm{sin }\ \lambda t\right),\\
z(t) = e^{-\Gamma t/2} \left( z_0 \mathrm{cos }\ \lambda t + \frac{ \Gamma z_0-2 \Omega x_0}{2 \lambda} \mathrm{sin }\ \lambda t\right), \label{eq:zens}
\end{eqnarray}
where $\lambda = \sqrt{\Omega^2-(\Gamma/2)^2}$.  We use (\ref{eq:zens}) to determine the Rabi frequency $\Omega$ for each measurement strength and Rabi drive amplitude.

\subsection{Propagation of the qubit state density matrix}

Given the Rabi frequency $\Omega$, the coherence decay rate $\gamma$, and the initial qubit state calculated from the values of $x_0$ and $z_0$ at time $t=0$, we propagate the initial state to states at later time steps $t = d t, 2 d t , ..., n d t$ using a two-step procedure.  At any time $t$, we first apply a unitary rotation to account for the Rabi drive, 
\begin{eqnarray}
\rho'_{01}= \rho_{01}+\frac{\Omega}{2} (\poo- \pii) d t, \label{eq-update1}\\
\rho'_{11} =\rho_{11} +\frac{\Omega}{2} (\rho'_{01}+\rho'_{10}) d t,
\end{eqnarray}
where $\rho_{00},\rho_{01},\rho_{10},\rho_{11}$ are matrix elements of a qubit density matrix $\rho(t)$. With the input values $\rho'_{01}, \rho'_{11}$, we next apply the Bayesian update to them based on the measurement result $V(t)$ obtained in the time interval between $t$ and $t+dt$ and get,
\begin{eqnarray}
\rho_{11}(t+dt) = \frac{(\pii'/\poo')\exp(-4 V(t) dt/ \tau\Delta V)}{1+(\pii'/\poo')\exp(-4 V(t) dt/\tau \Delta V)} ,\\
\rho_{01}(t+dt)=  \poi' \frac{\sqrt{(1-\pii(t+dt))\pii(t+dt)}}{\sqrt{(1-\pii')\pii'}} e^{-\gamma dt}. \label{eq-update2}
\end{eqnarray}
We use $dt = 16$ ns as the data sampling interval, and $V(t)$ is the measurement result obtained between $t$ and $t+dt$.  As discussed in the main text, we validate the state update procedure using conditioned quantum state tomography and find good agreement between individual trajectories and the tomographic reconstructions.

Moreover, in the time-continuum limit $dt \rightarrow 0$, we can approximate the state update procedure Eqs.~\eqref{eq-update1}-\eqref{eq-update2} with the differential equations,
\begin{align}
\dot{x}(t) =& - \gamma\, x(t)+ \Omega\, z(t) - x(t) z(t) r(t) / \tau,\label{eq-xz1} \\
\dot{z}(t)  =&  \,\,-\Omega\, x(t) + (1-z(t)^2) r(t) /\tau,\label{eq-xz2}
\end{align}
where $r(t) = 2 V(t)/\Delta V $ is the dimensionless measurement signal and $x(t) = \text{Tr}[\sigma_x \rho(t)] $, $z(t) = \text{Tr}[\sigma_z \rho(t)] $ are the Bloch sphere coordinates as functions of time.

%function rabiupdate(dumin,ohmegar)
%wave dumin
%variable ohmegar
%ohmegar*=2*pi
%variable deltat =.016
%variable snr = .203
%variable deltav= 4.11
%variable eta = .412
%variable gamt = snr/8*(1-eta)/eta+deltat/15 
%make/o/n=(dimsize(dumin,0)) rho11
%make/o/c/n=(dimsize(dumin,0)) rho01
%rho11[0] = .50
%rho01[0] = cmplx(0,.405)
%variable impt, rho11old
%variable i,k, re, im
%for(i=1;i<dimsize(dumin,0);i+=1)
%re = real(rho01[i-1])
%im = real(ohmegar*deltat/2*(2*rho11[i-1]-1))+ imag(rho01[i-1])
%rho01[i] = cmplx(re,im)
%impt = imag(rho01[i])
%rho11[i] = -ohmegar*deltat*impt+rho11[i-1]
%rho11old=rho11[i]
%rho11[i] = rho11[i]/(1-rho11[i])*exp(dumin[i-1]*snr/deltav)/(1+rho11[i]/(1-rho11[i])*exp(dumin[i-1]*snr/deltav))
%rho01[i] = rho01[i]*sqrt((1-rho11[i])*rho11[i])/sqrt((1-rho11old)*rho11old)*exp(-gamt)
%endfor
%make/o/n=(dimsize(dumin,0)) sigztraj, sigxtraj, sigytraj
%sigztraj = -(1-2*rho11)
%sigxtraj = 2*imag(rho01)
%sigytraj = 2*real(rho01)
%setscale/i x,0,dimsize(dumin,0)*.016,sigztraj, sigxtraj, sigytraj
%end

 \begin{figure*}
 \begin{center}
\includegraphics[width=.8\textwidth]{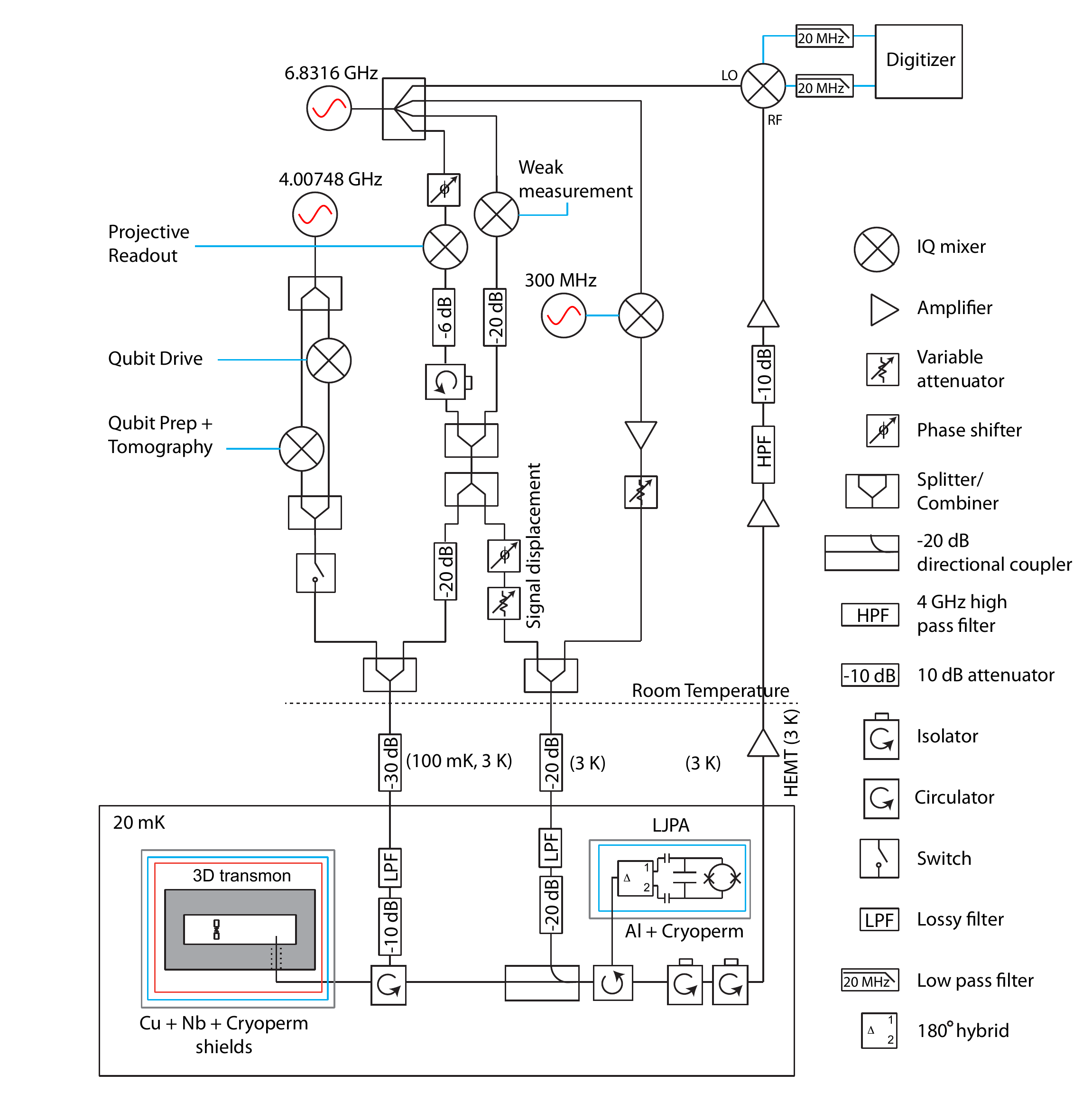}
\caption{\label{fig3} Experimental schematic.  The  weak measurement tone is always on. The projective readout tone is pulsed.  The amplitude and phase of the signal displacement tone are adjusted to displace the measurement signals back to the origin of the $X_1 X_2$ plane and allows the LJPA to perform in the linear regime.} 
\end{center}
\end{figure*}

%%%%%%%%%%%%%%%%%%%%%%%%%%%
% Theoretical methods
%%%%%%%%%%%%%%%%%%%%%%%%%%%

\section{Theoretical methods}

\subsection{Derivation of the ordinary differential equations Eq. (2) in the main text}

We consider a set of unitless measurement readouts $\{ r_k \} = \{r_0, r_1,...,r_{n-1}\}$ where $r_k = 2 V_k/\Delta V$ at times $\{t_k\}$ for $k=0,1,...,n-1$ and its corresponding set of qubit states denoted by $\{ {\bm q}_k \}$. In our experiment,  the $y$ component of the qubit Bloch coordinates is always zero, thus ${\bm q}_k$ is a 2-dimensional vector ${\bm q}_k = (x_k , z_k)$. We write a joint probability density function of all measurement outcomes $\{ r_k \}$, the quantum states $\{ \bm{q}_k \}$ and the chosen final state $\bm{q}_F$, conditioned on the initial state $\bm{q}_I$ as,
\begin{align}\label{eq-jpdf}
 \,\,P(\{ \bm{q}_k \},\{ r_k \},\bm{q}_F|\bm{q}_I)
&=\,\, \delta^2(\bm{q}_0-\bm{q}_I)\delta^2(\bm{q}_n-\bm{q}_F)
\left(\prod_{k=0}^{n-1}P(\bm{q}_{k+1}|\bm{q}_k,r_k)\, P(r_k|\bm{q}_k)\right).
\end{align}
Here, $P(\bm{q}_{k+1}|\bm{q}_k,r_k)$ is a probability density function of a qubit state at time $t_{k+1}$ given a qubit state and measurement signal at previous time $t_k$. Since a qubit state at any time $t_{k+1}$ is updated deterministically from ${\bm q}_k$ and $r_k$, the density function $P(\bm{q}_{k+1}|\bm{q}_k,r_k)$ is a delta function with the state update equations. The conditional distribution of the detector output $P(r_k|\bm{q}_k)$ is a probability density function of $r_k$ given ${\bm q}_k$ which is,

\begin{align}\label{eq-probr}
P(r_k | \bm{q}_k) =\sqrt{\frac{\delta t}{2 \pi \tau}}\left(\frac{1+z_k}{2} e^{-\frac{\delta t}{2 \tau}(r_k-1)^2} +\frac{1-z_k}{2} e^{-\frac{\delta t}{2 \tau}(r_k+1)^2}\right).
\end{align}

By expressing the delta functions in  Eq.~\eqref{eq-jpdf} in Fourier forms with conjugate variables ${\bm p}_k = ( p^x_k, p^y_k)$ for $k=-1,0,...,n$ and other terms in exponential forms, we can write the joint probability density function in a path integral representation $P(\{ \bm{q}_k \},\{ r_k \},\bm{q}_F|\bm{q}_I) \propto  \,\, \int\!{\cal D}{\bm p}\, e^{{\cal S}}$. Here ${\cal D}{\bm p}$ is an integral measure over conjugate variables $\{ {\bm p}_k \}$ and ${\cal S}$ is an action of the integrals given by,
\begin{subequations}\label{eq-action}
\begin{align}
\label{eq-action1} {\cal S}  = & - \bm{p}_{-1} \cdot (\bm{q}_0-\bm{q}_I) - \bm{p}_n \cdot (\bm{q}_n-\bm{q}_F) + \sum_{k=0}^{n-1} \bigg\{- \bm{p}_k \cdot (\bm{q}_{k+1}-\bm{\mathcal{E}}[\bm{q}_k,r_k]) + \ln P(r_k | \bm{q}_k) \bigg\},\\
\label{eq-action2}=&- B + \int_0^T\!\!\! {\rm d} t \big[ -  p_x \dot{x}-  p_z \dot{z} + p_x(-\gamma x + \Omega z-x z r/\tau) + p_z (-\Omega x + (1-z^2)r/\tau) -(r^2- 2 r z + 1)/2 \tau\big],
\end{align}
\end{subequations}
where we have used the operator ${\bm{\mathcal E}}[\bm{q}_k,r_k]$ for the state update equations and $B$ as a short hand for the first two terms in Eq.~\eqref{eq-action1}. We note that, in the second line, we have taken the time-continuum limit $\delta t \rightarrow 0$ and written the action explicitly for our qubit measurement case with the state update equations  Eq.~\eqref{eq-xz1}-\eqref{eq-xz2}. We have also used shortened notation of variables, e.g., $x = x(t) \equiv \lim_{\delta t \rightarrow 0} \{ x_0, x_1, ..., x_n\}$. To obtain the most likely path, we then extremize the action Eq.~\eqref{eq-action2} over all variables $x,z,p_x,p_z, r$ and obtain the ordinary differential equations as shown in Eq.~(2) in the main text,
\begin{subequations}\label{eq-odes}
\begin{eqnarray}  
\dot{x} &=& -\, \gamma \, x  + \Omega\, z - x z\, r /\tau,  \\
\dot{z} & =& -\, \Omega \, x + (1-z^2)\, r /\tau,  \\
\dot{p}_x &=& +\,\gamma \, p_x +\Omega\, p_z + p_x z \,r/\tau, \label{eq-odespx} \\ 
\dot{p}_z &=& -\,\Omega\, p_x + (p_x x + 2 p_z z -1)\,r/\tau, \label{eq-odespz}
\end{eqnarray}
\end{subequations}
where $r =\, z +   p_z (1-z^2) -  p_x x z $ and the forced boundary conditions are $x(t=0) = x_I,  z(t=0) = z_I , x(t=T) = x_F , z(t=T) = z_F$. As discussed in the main text, we can analytically solve the ODEs Eqs.~\eqref{eq-odes} when $\Omega =0$. For the driven case where $\Omega \ne 0$, we solve the equations numerically using a shooting method.

%As discussed in the main text, we can analytically solve the ODEs Eqs.~\eqref{eq-odes} when $\Omega =0$. However, for the driven case where $\Omega \ne 0$, we solve the equations numerically using a shooting method. In some cases, we can find multiple solutions by changing the initial values of $x, z, p_x, p_z$ to start the shooting process. We discuss the multiple solutions in more detail in subsection~\ref{sec-multi}

\subsection{Interpretation of the solutions of the ODEs}\label{sec-pathprob}

Here we discuss the interpretation of the solution of the ODEs Eq.~\eqref{eq-odes} (also Eq.~(2) in the main text). The extremization of the action Eq.~\eqref{eq-action1} can also be interpreted as a constrained optimization of the last term of Eq.~\eqref{eq-action1}, $\sum_{k=0}^{n-1}\ln P(r_k | \bm{q}_k)$, the log-likelihood of the trajectory. The constraints are, 1)~the qubit state updates $\bm{q}_{k+1} = \bm{\mathcal{E}}[\bm{q}_k,r_k]$ for $k=0,1,...,n-1$, 2)~the pre-selected state ${\bm q}_0 = {\bm q}_I$, and 3)~the post-selected state ${\bm q}_n = {\bm q}_F$. The conjugate variables $\{ {\bm p}_k \}$ are now act as the Lagrange multipliers of the constrained optimization. With this interpretation, a solution of the ODEs Eq.~\eqref{eq-odes}, therefore, represents a path with an optimized value of $\sum_{k=0}^{n-1}\ln P(r_k | \bm{q}_k)$ or its exponential $\prod_{k=0}^{n-1} P(r_k |\bm{q}_k)$, i.e., a measurement path probability density.

\subsection{The most likely path}

The optimized path mentioned in the previous subsection \ref{sec-pathprob} can represent either a maximum, a minimum or a saddle point of the path probability under the constraints. We can determine this by finding paths slightly varied from the optimized solution, with all constraints still applied. This can be done by adding small constants $\delta_1, \delta_2$  to the right hand side of the differential equations of the conjugate variables $p_x, p_z$ Eqs.~\eqref{eq-odespx},\eqref{eq-odespz}, leaving the equations of $x,z$ unchanged, and solving them with the same boundary conditions. Solutions of this modified ODEs will be slightly varied from the optimized path. We, then, compute their full-path probabilities, comparing with the probability calculated from the optimized path. In Fig.~\ref{fig-MLPvariation}, we show samples of paths from the variational method described here and the unnormalized full-path probability of the surrounding paths. In this case, it shows that the optimized solution is the most likely path with a maximum value of the path probability density.

\begin{figure*}
\includegraphics[width=18cm]{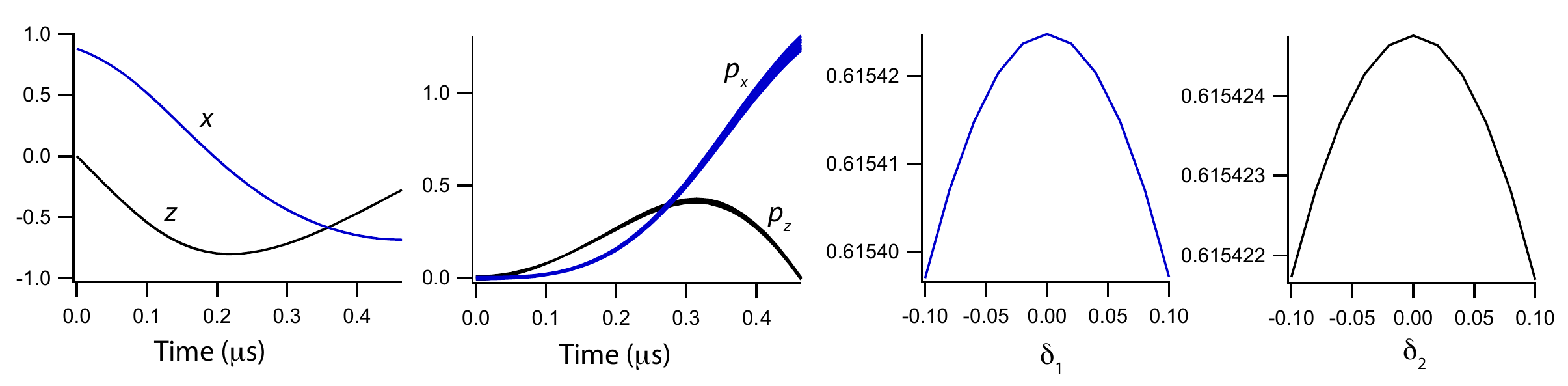}
\caption{From left to right, (1) $x$ and $z$ coordinates of 11 trajectories slightly varied from an optimized solution with boundary conditions $ (x_I, z_I) = (0.88,0)$, $(x_F, z_F, T_F) = (-0.683,-0.227, 0.464 \mu s)$ and the Rabi drive $\Omega/2 \pi = 1.08$ MHz, (2) corresponding conjugate variables $p_x$ and $p_z$, (3) a plot of the unnormalized probability versus changes of a constant $\delta_1$ in the $p_x$ differential equation and (4) unnormalized probability versus changes of a constant $\delta_2$ in the $p_z$ differential equation. In this case, the optimized solution gives a maximum value of the path probability density.}\label{fig-MLPvariation}
\end{figure*}

\subsection{The most likely paths from the experimental post-selected trajectories}\label{sec-numer}

To find the most likely path from experimental trajectory data, we post-select trajectories starting from the same initial state ${\bm q}_I = ( x_I, z_I)$ and ending around a final state ${\bm q}_F = ( x_F, z_F)$ with a small tolerance. Then, we compute the path probability $\prod_{k=0}^{n-1} P(r_k |\bm{q}_k){\rm d}r_k$ described in subsection \ref{sec-pathprob} using Eq.~\eqref{eq-probr} for each trajectory in the post-selected sub-ensemble. (In principle, we can also compute the path probability directly from the frequencies of the trajectories, however, the statistical convergence of the latter method is much slower.) This path probability indicates the relative likelihood for trajectories to be in a tube of volume ${\rm d}v={\rm d}r_0{\rm d}r_1\cdots {\rm d}r_{n-1}$ around the given path. We choose the top $\sim 5$ - $8\%$ of the post-selected trajectories that give the largest values of the path probability and average them to obtain an approximation to the most likely path. We show in Fig.~\ref{fig-correctedMLpath} the paths with the same sets of post-selection conditions as used in Figure 4 of the main text. The experimental most likely paths, even with their jaggedness, closely approximate the theoretical most likely paths, solutions of the ODEs Eqs.~\eqref{eq-odes}. We expect that the approximated curves should converge to the smooth theory curves in the limit of an infinite ensemble of the post-selected trajectories.
\\

\begin{figure*}
\includegraphics[width=15cm]{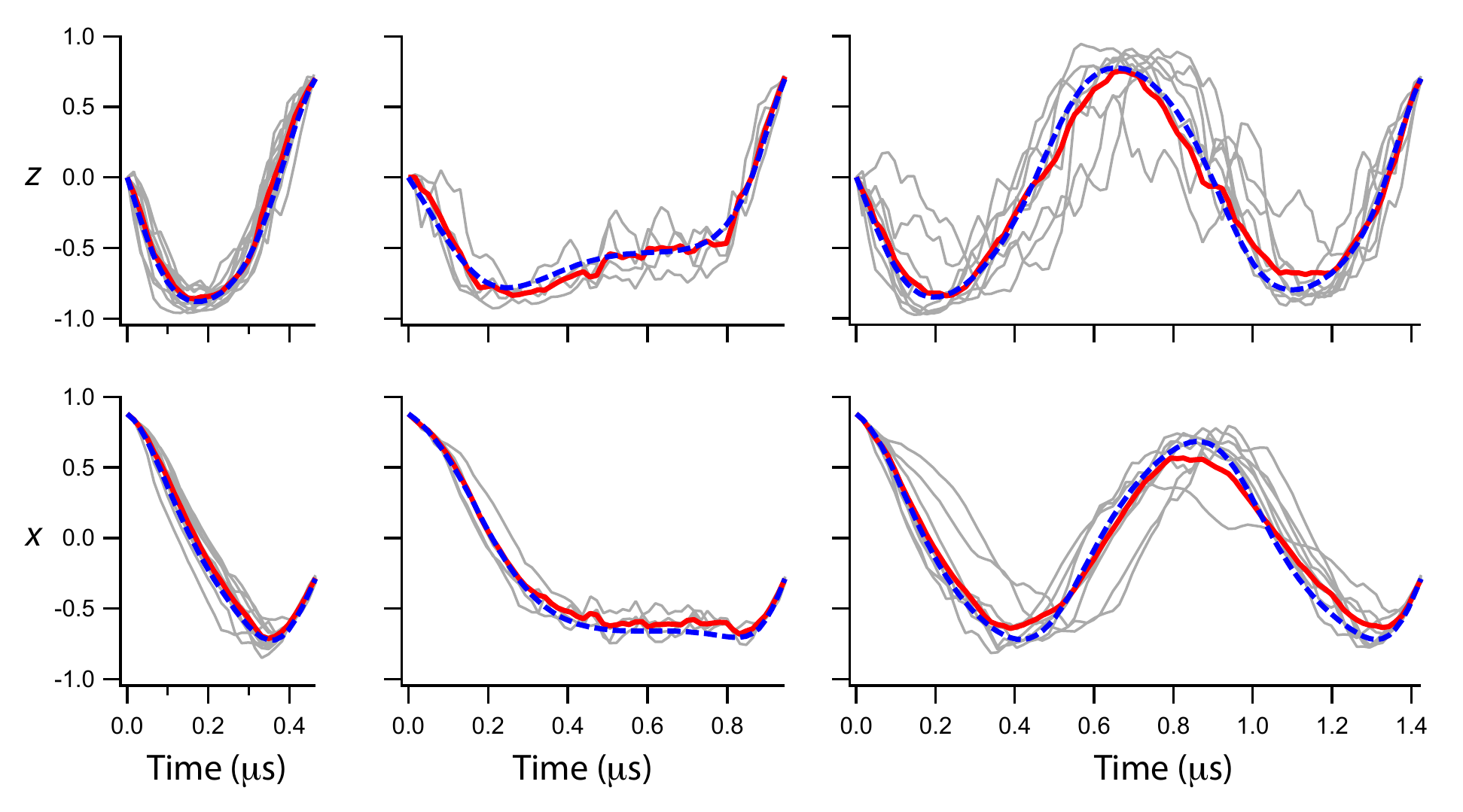}
\caption{With the same sets of post-selections condition as used in Figure 4 of the main text, we show the theoretical most likely paths (dashed blue), the top $\sim 5$ - $8\%$ of the post-selected trajectories that give the largest values of the path probability (grey), and their average (red). The average (red) curves are the approximation of the most likely paths as described in subsection~\ref{sec-numer}. The top row is for $z$ coordinate and the bottom row is for $x$ coordinate. The post-selection conditions are (from left to right) : $(x_F, z_F, T_F) = (-0.29, 0.7, 0.464 \mu s), (-0.29, 0.7, 0.944 \mu s), (-0.29, 0.7, 1.424 \mu s)$ with post-selection tolerance 0.03.}\label{fig-correctedMLpath}
\end{figure*}

In some cases, we can simply look at a trajectory of local medians (medians of $x$ or $z$ at all time steps) and compare it with the theoretical most likely path. The median trajectory can practically be a good approximation to the theory curve when the distribution of the post-selected trajectories is a narrow band, i.e., the post-selected trajectories lie closely around a single path. As an example, in the case where there is no drive on the qubit, $\Omega = 0$, we show in our theory paper (PRA 88, 042110) that the median curves agree quite well with the most likely curves, solutions of the ODEs. However, in the driven case where the qubit trajectories can possibly have different winding numbers around the $y$-axis resulting in multiple probable paths from an initial state to a final state, simply finding the medians of the distribution of $x$ or $z$ is not enough to capture their most likely behaviour. In this paper, we only focus on the cases where there is a single most likely path between any two boundary states. We will discuss our findings concerning the multiple paths connecting two boundary states in a future work.

\subsection{The most likely time}

\begin{figure}
\includegraphics[width=9cm]{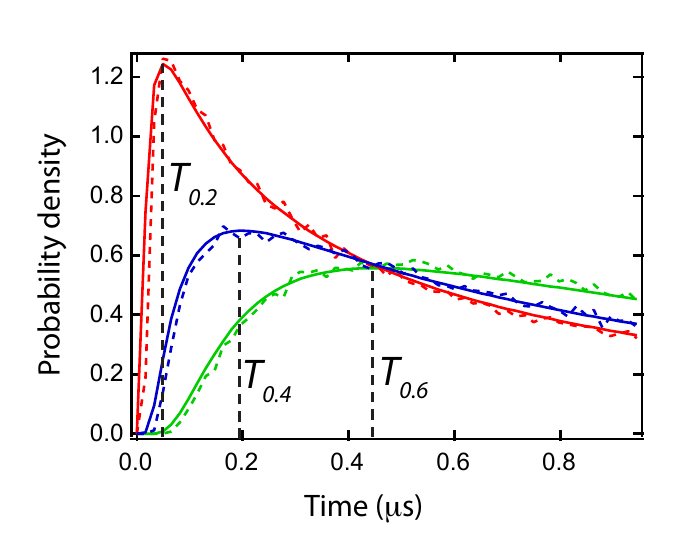}
\caption{The probability density functions $P(z_F|z_I=0)$ are plotted as a function of time $T$  (solid curves) along with experimental data  (dotted curves) with $\tau = 1.25 \mu$s. The red, green, and blue curves are the distribution functions $P(z_F=0.2|z_I=0)$, $P(z_F=0.4|z_I=0)$, and $P(z_F=0.6|z_I=0)$, respectively. The optimized times $T_{\text{opt}}$ for the three cases are shown as the vertical black dashed lines with the labels $T_{0.2}, T_{0.4}, T_{0.6}$.}
\label{fig-newtopt}
\end{figure}

Besides the path of maximum likelihood taken between the pre- and post-selected states in a fixed time, a complementary problem in quantum control is that of the optimal waiting time between starting and destination states. In the case where there is no Rabi drive on the qubit, $\Omega =0$, we can fix the states at the endpoints and inquire about the most likely time taken to travel between them. While a path-integral derivation of the most likely time is possible, we give a simpler derivation here based on the probability distribution of the time-average measurement readout $V = (1/n)\sum_{k=0}^{n-1} V_k$. 

 In the case with no drive on the qubit, the $z$-coordinate of the qubit on the Bloch sphere at any time $T$ is solely determined by the time-average measurement readout $V$. We can derive the distribution of the final $z$-coordinate ($z_F$) at any time $T$ given the initial $z$-coordinate ($z_I$), $P(z_F | z_I)$, from the probability density function $P(V|z_I)$. The probability density function of the average measurement outcome $V$ given the initial qubit's $z$-coordinate $z_I$ is,
\begin{align}\nonumber
P(V \,|\, z_I )= &\, P(V \, |\, 0) \frac{1+z_I}{2} + P(V \,|\, 1) \frac{1-z_I}{2}\\
=&\,\sqrt{\frac{ 1}{2 \pi \sigma^2}}\,\bigg( \frac{1+z_I}{2} e^{-\frac{ 1}{2 \sigma^2}(V-\Delta V/2)^2}+ \frac{1-z_I}{2}e^{-\frac{ 1}{ 2 \sigma^2}(V+\Delta V/2)^2}\bigg),
\end{align}
where the variance of the transmon voltage signal measured in $\delta t$ duration is $\sigma^2 = \Delta V^2 \,\tau/ 4 \, \delta t$. We change variables from the time-averaged measurement signal $V$ to the final $z$-component $z_F$ by $V  =  \frac{\tau \Delta V}{2T}(\tanh^{-1} z_F - \tanh^{-1} z_I)$. We obtain the differential measure ${\rm d}V = \frac{\tau}{ 2T}\frac{ \Delta V}{ (1-z_F^2)}{\rm d}z_F$. The probability density function of $z_F$ given $z_I$ can be computed via a relation $P(V \,|\, z_I ){\rm d}V = P(z_F \,|\,z_I) {\rm d}z_F$,
\begin{align}\label{eq-probzf}
P(z_F|z_I) = \frac{\sqrt{\frac{\tau}{2 \pi T}}}{(1-z_F^2)} \exp\left\{ -\frac{T}{2 \tau}(\bar{r}^2+1) +\frac{1}{2} \ln \big(\frac{1-z_I^2}{1-z_F^2}\big)\right\},
\end{align}
where $\bar{r} \equiv \frac{\tau}{T} \tanh^{-1}\big(\frac{z_F-z_I}{1-z_I z_F}\big) = \frac{\tau}{T}(\tanh^{-1} z_F - \tanh^{-1} z_I)$. For the case where the initial state is $x = +1$,  $(z_I=0)$, the probability density function simplifies to
\begin{align}
P(z_F|z_I=0) = \frac{\sqrt{\frac{\tau}{2 \pi T}}}{(1-z_F^2)^{\frac{3}{2}}}\exp\left\{ -\frac{T}{2 \tau}-\frac{\tau}{2 T}(\tanh^{-1} \!z_F)^2 \right\}.
\end{align}

We then compute the most likely time $T_{\text{opt}}$ where the probability density function $P(z_F | z_I)$ is maximized for the fixed values of $z_I$ and $z_F$. By maximizing the probability function $P(z_F | z_I)$ with respect to $T$, we obtain,
\begin{align}\label{eq-timeoptright}
T_{\text{opt}} = \tau\left(\frac{\sqrt{1+ 4 \,\bar{\gamma}^2}-1}{2}\right),
\end{align}
where $\bar{\gamma} \equiv \tanh^{-1}\big(\frac{z_F-z_I}{1-z_I z_F}\big)$. We show in Fig.~\ref{fig-newtopt} the distributions $P(z_F|z_I=0)$ as a function of time $T$ for $z_F = 0.2, 0.4, 0.6$. They show very good agreement with the experimental data.

\section{Extended results}

Here, in Figure~\ref{fig:gallery}, we display an extended set of experimental results for different driving and measurement parameters. The histograms of all quantum trajectories as well as the post-selected trajectories are shown for different Rabi drives, measurement strengths, and post-selected states at different times.

\begin{figure*}
\includegraphics[width=0.8\textwidth]{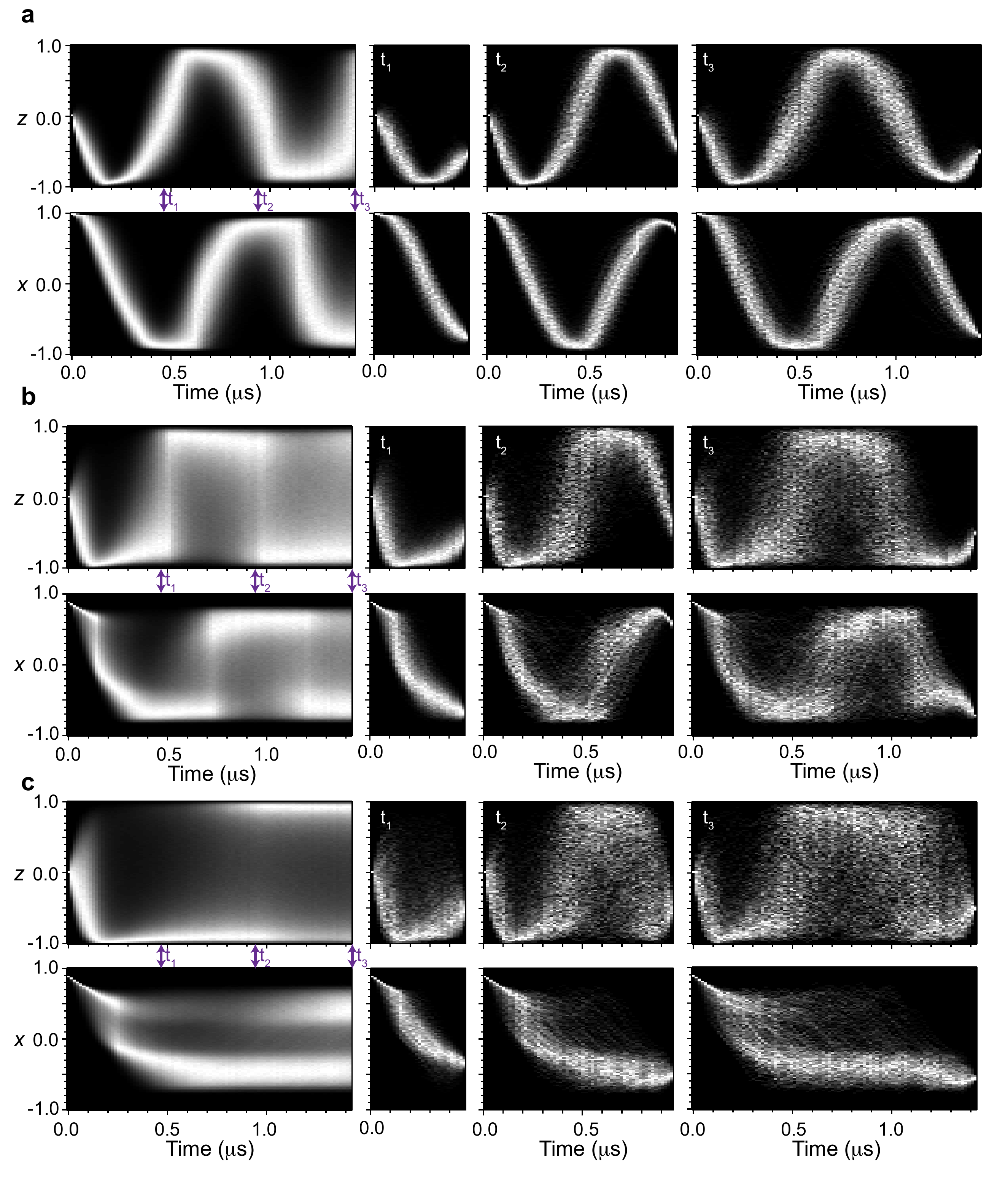}
\caption{Greyscale histograms of ensemble and post-selected trajectories for different Rabi frequencies and measurement strengths.  {\bf a} Ensemble and post-selected trajectories for $\Omega/2\pi = 1.08$ MHz and $\tau = 1.25 \ \mu$s.  The post-selections for times $\{t_1 = 464$ ns, $t_2= 944$ ns, $t_3 = 1.424 \ \mu$s$\}$ are $(x_F,z_F) = \{(-0.78,-.5), (0.7,-.5), (-0.73,-.5)\}$ with a post selection window of $\pm 0.03$. {\bf b} Trajectories for $\Omega/2\pi = 1.08$ MHz and $\tau = 315 $ ns with $(x_F,z_F) = \{(-0.69,-.5), (0.5,-.5), (-0.73,-.5)\}$.  {\bf c} Trajectories for $\Omega/2\pi = 0.58$ MHz and $\tau = 315 $ ns with $(x_F,z_F) = \{(-0.35,-.5), (-0.5,-.5), (-0.56,-.5)\}$. Note that all the trajectories use the same value of $z_F$.  The values of $x_F$ were chosen to give a large number of trajectories in the post-selected ensemble. }

\label{fig:gallery}
\end{figure*}

\end{widetext}

\end{document}